\documentstyle[epsf]{elsart}
\begin{document}

\def\note#1{{\bf[#1]}}

\def\dirac{{\bf \rm D}\!\!\!\!/\,}
\def\wilson{{\bf \rm W}}
\def\ham{{\bf \rm H}}
\def\mbham{{\cal H}}
\def\bmat{{\bf \rm B}}
\def\cmat{{\bf \rm C}}

\begin{frontmatter}

\begin{flushright}
{\normalsize FSU-SCRI-98-61}\\
\end{flushright}

\title{ 
Evidence for fractional topological charge in SU(2)
pure Yang-Mills theory
}
\author{
Robert G. Edwards, Urs M. Heller and Rajamani Narayanan}
\address{
SCRI, The Florida State University, 
Tallahassee, FL 32306-4130, USA}

\begin{abstract}
We investigate the spectral flows of the hermitian Wilson-Dirac
operator in the fundamental and adjoint representations on two
ensembles of pure SU(2) gauge field configurations at the same
physical volume.  We find several background gauge field
configurations where the index of the hermitian Wilson-Dirac operator
in the adjoint representation is not four times the index in the
fundamental representation.  This could imply a topological basis for
the gluino condensate in supersymmetric Yang-Mills theories.
\end{abstract}

\end{frontmatter}

{\bf PACS \#:}  11.15.Ha, 11.30.Pb\hfill\break

The index of the massless Dirac operator in the adjoint representation
of the SU(N) gauge group in a background field of topological charge
$Q$ is equal to $2NQ$~\cite{ad_index}.  Classical instantons carry an
integer topological charge and therefore can only cause a condensation
of an operator that contains $2N$ Majorana fermions. But it has been
argued on very general grounds that there is a bilinear gluino
condensate in supersymmetric Yang-Mills theories~\cite{Witten}.
Self-dual twisted gauge field configurations with a fractional
topological charge of $1\over N$~\cite{Hooft} can be used to explain a
bilinear gluonic condensate~\cite{Cohen}.  A non-perturbative
computation of the gluonic condensate on the lattice would typically
be done with conventional periodic boundary conditions for both the
gauge fields and fermions in order to achieve a supersymmetric theory
in the continuum limit.  In this situation it is not clear as to what
causes the bilinear gluonic condensate. In addition, a condensation is
expected in a finite physical volume where the spectrum is discrete.

The overlap formalism for chiral fermions~\cite{over} provides a
definition of the index of the chiral Dirac operator on the lattice.
In this paper we use this definition to measure the index of the
chiral Dirac operator in the adjoint representation in an ensemble of
pure SU(2) gauge field 
backgrounds.\footnote{The relationship between the gluino condensate and gauge
field topology in the overlap formalism was addressed in Ref.~\cite{index}.}
Since the fermion is in the real
representation of the gauge group, the spectrum of the hermitian
Wilson Dirac operator occurring in the overlap formalism is doubly
degenerate. Therefore, the index of the associated chiral Dirac
operator  defined by the overlap
formalism can only be even valued. This is to be expected. Otherwise,
we would have to explain away unphysical expectation values of an odd
number of fermionic observables in an otherwise well defined theory.
However, it is possible to obtain any even value for the index.  The issue
one has to address is if all possible even values are realized for the
index in an ensemble of SU(2) gauge field backgrounds or if only
multiples of four are observed. If only multiples of four are observed
then one would conclude that the gauge field background behaves as if
it were made up of classical instantons with small fluctuations, and we
would not be able to explain the bilinear condensate by topological
means. On the other hand, if we observe any even value for the index
that are not just multiples of four, the background gauge fields
cannot be thought of as being made up of classical instantons and we
would be able to explain the bilinear condensate through topological
means.

In this paper we consider two ensembles of SU(2) background gauge
fields generated using the standard Wilson action with periodic
boundary conditions on the gauge fields.  One is at a lattice coupling
of $\beta=2.4$ on an $8^4$ lattice and the other is at a lattice
coupling of $\beta=2.6$ on a $16^4$ lattice. The couplings were chosen
so that both lattices had the same physical volume ($a=0.12$fm at
$\beta=2.4$ and $a=0.06$fm at $\beta=2.6$) and is in the confined
phase ($\beta_c=2.5115$ at $N_\tau=8$ and $\beta_c=2.74$ at
$N_\tau=16$)~\cite{Heller}.  Each ensemble contains a total of fifty
configurations.  We have doubled this number by including the parity
transformed partner of every configuration since this symmetrizes the
distribution of the index.  Using the overlap definition of the index,
we compute the index of the chiral Dirac operator in the fundamental
and adjoint representations.  We refer to these quantities as $I_f$
and $I_a$, respectively. We do not present any details about the
definition of the index using the overlap formalism and the technical
details involved in the numerical computation of the index. The
definition of the index in the overlap formalism can be found in
Ref.~\cite{over}. We keep track of levels crossing zero to measure the
index which involves a computation of the low lying modes of the
hermitian Wilson-Dirac operator~\cite{su3_top}.  The results from the
measurements of $I_f$ and $I_a$ are tabulated in 
Tables~\ref{tab:8}
and 2
for the two ensembles.  Each column is associated with
a fixed value of $I_f$ and each row is associated with a fixed value
of $I_a$.  We find that $I_a$ comes only in multiples of two as
explained in the previous paragraph.  The numbers in the second row
are the number of configurations with a fixed $I_f$.  The numbers in
the second column are the number of configurations with a fixed
$I_a$. All other entries in the table tell us the number of
configurations with a fixed value of $I_f$ and $I_a$.  The numbers in
bold are the configurations where $I_a=4I_f$.  Clearly, there are
several configurations in both ensembles that do not satisfy the
relation $I_a=4I_f$, and we find several configurations where $I_a$ is
not a multiple of four.  The occurrence of a significant number of
configurations with values of $I_a$ that are not multiples of four is
taken as evidence for the existence of gauge field configurations with
fractional topological charge in our ensemble since $Q={I_a\over 4}$
is the continuum relation between the topological charge and the index
of the Dirac operator in the adjoint representation. 

\begin{figure}
\epsfxsize=7in
\centerline{\epsffile{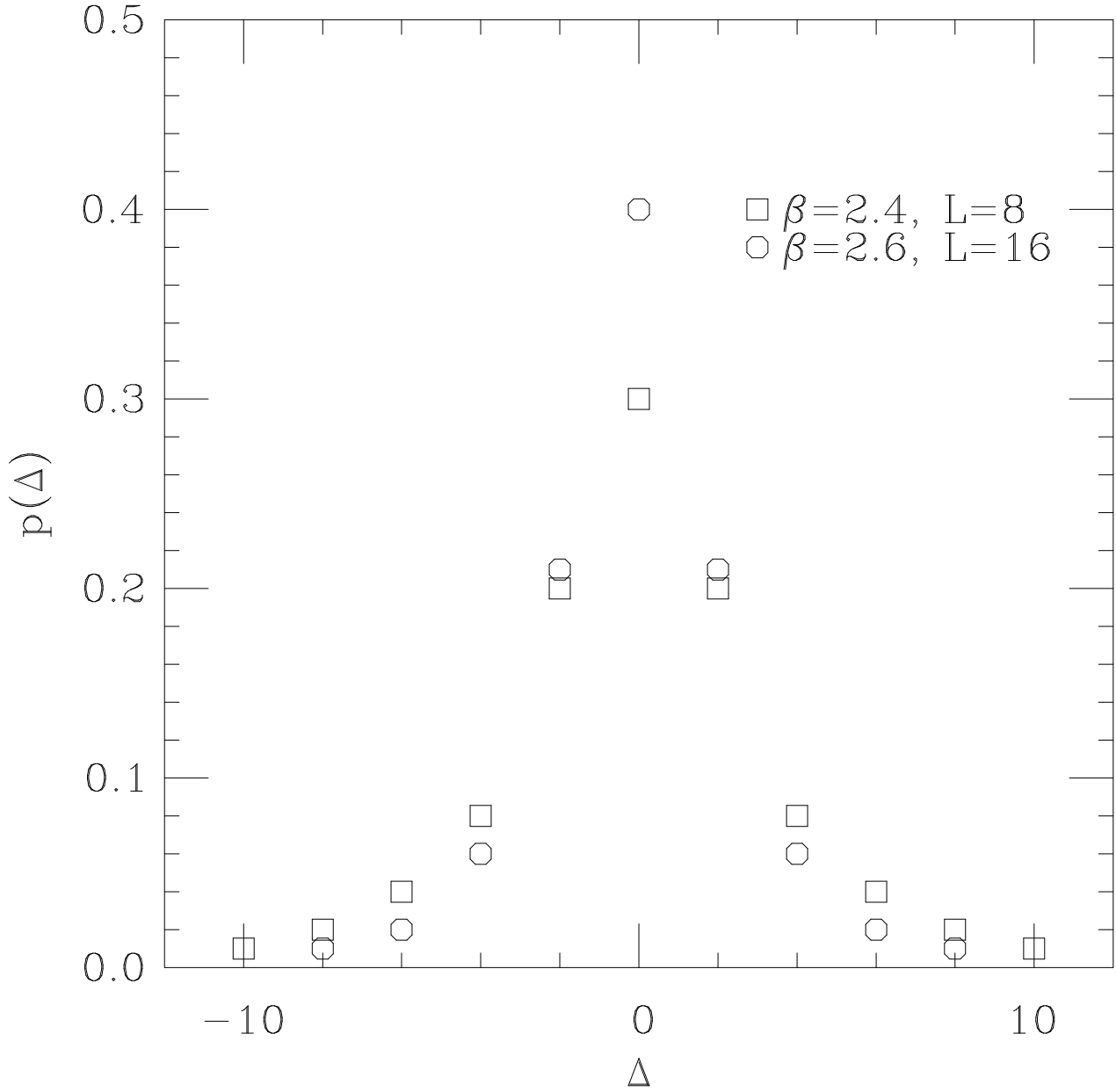}}
\caption{
$p(\Delta)$ versus $\Delta$ for the two ensembles where $\Delta=I_a-4I_f$.
}
\label{fig:delta}
\end{figure}

Having provided some evidence for the existence of fractional
topological charge on the lattice at finite lattice spacing, we now
address the question of whether these are pure lattice artifacts.  For
this purpose, we define the quantity $\Delta=I_a-4I_f$ for each
configuration in both ensembles.  Note that $\Delta$ takes on only
even values.  The probability of finding a certain value of $\Delta$,
$p(\Delta)$, is plotted for the two ensembles in
Fig.~\ref{fig:delta}. We find that $p(\Delta)$ for $|\Delta| > 2$
decreases as one goes toward the continuum limit at a fixed physical
volume.  However, $p(\pm 2)$ does not decrease indicating that it
might remain finite in the continuum limit.

In this paper we have presented some preliminary evidence for
fractional topological charge on the lattice. By studying two
ensembles with different lattice spacings we have argued that there is
a reasonable indication that this is not a lattice artifact.  If this
result survives the continuum limit it provides a topological basis
for the gluino condensate in supersymmetric gauge theories.

\ack{
We would like to thank Herbert Neuberger and Mark Alford for discussions.
This research was supported by DOE contracts 
DE-FG05-85ER250000 and DE-FG05-96ER40979.
Computations were performed on the QCDSP and CM-2 at SCRI.}

\vfill\eject

\begin{table}[t]
\begin{center}
\begin{tabular}{|c||c||c|c|c|c|c|} \hline
& $I_f$    & -2   & -1    & 0      & 1     & 2  \\ \hline\hline
$I_a$ & \# &  10  & 20    &  40    & 20    & 10 \\ \hline\hline
-6    & 2  &  1   & 1     &        &       &    \\ \hline
-4    & 10 &  4   &{\bf 4}&  2     &       &    \\ \hline
-2    & 23 &  2   & 11    &  7     & 2     & 1  \\ \hline
 0    & 30 &  2   & 2     &{\bf 22}& 2     & 2  \\ \hline
 2    & 23 &  1   & 2     &  7     & 11    & 2  \\ \hline
 4    & 10 &      &       &   2    &{\bf 4}& 4  \\ \hline
 6    & 2  &      &       &        & 1     & 1  \\ \hline
\end{tabular}
\caption{
Results from the measurement of the index of the chiral Dirac
operator in the fundamental and adjoint representations at $\beta=2.4$ on an
$8^4$ lattice.}
\end{center}
\label{tab:8}
\end{table}

\begin{table}[t]
\begin{center}
\begin{tabular}{|c||c||c|c|c|c|c|c|c|c|c|} \hline
& $I_f$    & -4  & -3 & -2    & -1     & 0      & 1      & 2     & 3 & 4 \\ \hline\hline
$I_a$ & \# &  1  & 2  &  5    & 24     & 36     & 24     & 5     & 2 & 1 \\ \hline\hline
-12 & 1    &  1  &    &       &        &        &        &       &   &   \\ \hline
-10 & 1    &     & 1  &       &        &        &        &       &   &   \\ \hline
-8  & 1    &     &    &{\bf 1}&        &        &        &       &   &   \\ \hline
-6  & 6    &     & 1  & 3     & 2      &        &        &       &   &   \\ \hline
-4  & 12   &     &    &       &{\bf 10}& 2      &        &       &   &   \\ \hline
-2  & 16   &     &    &       & 8      & 7      & 1      &       &   &   \\ \hline
 0  & 26   &     &    & 1     & 3      &{\bf 18}& 3      & 1     &   &   \\ \hline
 2  & 16   &     &    &       & 1      & 7      & 8      &       &   &   \\ \hline
 4  & 12   &     &    &       &        & 2      &{\bf 10}&       &   &   \\ \hline
 6  & 6    &     &    &       &        &        & 2      & 3     & 1 &   \\ \hline
 8  & 1    &     &    &       &        &        &        &{\bf 1}&   &   \\ \hline
 10 & 1    &     &    &       &        &        &        &       & 1 &   \\ \hline
 12 & 1    &     &    &       &        &        &        &       &   & 1 \\ \hline
\end{tabular}
\caption{
Results from the measurement of the index of the chiral Dirac
operator in the fundamental and adjoint representations at $\beta=2.6$ on a
$16^4$ lattice.}
\end{center}
\label{tab:16}
\end{table}

\end{document}